\documentclass[pra,twocolumn,showpacs,superscriptaddress,floatfix]{revtex4}
\usepackage{epsfig}
\usepackage{latexsym}
\usepackage{amsmath}
\usepackage{graphics}
\usepackage{bm}

\newcommand\br{\mathbf{r}}

\newcommand\bk{\mathbf{k}}

\newcommand\romand{\mathrm{d}}

\begin{document}

\title{Vortex in a trapped
Bose-Einstein condensate with dipole-dipole interactions}
\author{Duncan H J O'Dell}
\affiliation{Centre for Cold Matter, The Blackett Laboratory, Imperial College, London, SW7 2BW, United Kingdom}
\author{Claudia Eberlein}
\affiliation{Dept of Physics \& Astronomy,
    University of Sussex,
    Falmer, Brighton BN1 9QH, United Kingdom}

\date{\today}
\begin{abstract}
We calculate the critical rotation frequency at which a vortex state becomes energetically favorable over the vortex-free ground state in a harmonically trapped Bose-Einstein condensate  whose atoms have dipole-dipole interactions as well as the usual $s$-wave contact interactions. 
In the Thomas-Fermi (hydrodynamic) regime, dipolar condensates in oblate cylindrical traps (with the dipoles aligned along the axis of symmetry of the trap) tend to have lower critical rotation frequencies than their purely $s$-wave contact interaction counterparts. The converse is true for dipolar condensates in prolate traps.  
Quadrupole excitations and centre of mass motion are also briefly discussed as possible competing mechanisms to a vortex as means by which superfluids with partially attractive interactions might carry angular momentum. 
\end{abstract}

\pacs{03.75.Lm, 34.20.Cf, 32.10.Dk, 75.80.+q}

 \maketitle

\section{Introduction}
The achievement of Bose-Einstein condensation (BEC) in a trapped gas of
$^{52}$Cr atoms by the Stuttgart group \cite{griesmaier05} is the first
instance of a condensate with large dipole-dipole interactions. In
comparison to alkali atoms, which have a maximum magnetic dipole moment of
$\mu=1\mu_{B}$ ($\mu_{B}$ is the Bohr magneton), chromium atoms possess an
anomalously large magnetic dipole moment of $\mu=6\mu_{B}$. The long-range
part of the interaction between two dipoles separated by $\br$, and aligned
by an external field along a unit vector $\hat{\mathbf{e}}$ (in this paper
we shall always assume the dipoles are aligned by an external field), is
given by
\begin{equation}
U_{\mathrm{dd}}(\br)= \frac{C_{\mathrm{dd}}}{4 \pi}\,
\hat{{\rm e}}_{i} \hat{{\rm e}}_{j} \frac{\left(\delta_{i
j}- 3 \hat{r}_{i} \hat{r}_{j}\right)}{r^{3}}
\label{eq:staticdipdip}
\end{equation}
where the coupling $C_{\mathrm{dd}}=\mu_{0} \mu^{2}$ depends on the square
of the dipole moment.  Chromium atoms also have shorter range isotropic
interactions which are asymptotically of the van der Waals-type and so fall
off as $r^{-6}$. At ultra-low temperatures the de Broglie wavelength of the
atoms is much larger than the range of the isotropic interactions which can
consequently be handled within perturbation theory by using the usual
delta-function pseudo-potential \cite{pitaevskii+stringari}
\begin{equation} U(\br)=4 \pi \hbar^{2} a_{\mathrm{s}}
\delta(\br) /m \equiv g \delta(\br)
\label{eq:deltafninteraction}
\end{equation}
characterized solely by the $s$-wave scattering length $a_s$. 
A measure of the
strength of the long-range dipole-dipole interaction relative to
the s-wave scattering energy is given by the dimensionless
quantity
\begin{equation}
\varepsilon_{\mathrm{dd}}\equiv \frac{C_{\mathrm{dd}}}{3 g}.
\label{eq:edd_definition}
\end{equation}
For $^{87}$Rb one finds $\varepsilon_{\mathrm{dd}} \approx 0.007$, and Na
has $\varepsilon_{\mathrm{dd}} \approx 0.004$. Taking the $s$-wave
scattering length of $^{52}$Cr to be $a_{s}=105a_{\mathrm{B}}$ \cite{griesmaier05}, one
obtains $\varepsilon_{\mathrm{dd}} \approx 0.144$. In a further significant
development, the Stuttgart group has also studied Feshbach scattering
resonances between $^{52}$Cr atoms \cite{werner05}. This means that the
magnitude and sign of $a_s$, and thus also the value of
$\varepsilon_{\mathrm{dd}} $, can be controlled.

The principal effect of the anisotropy of the dipole-dipole interactions
upon a stationary, trapped, condensate will be to distort its aspect ratio
so that it is elongated along the direction of the external field
\cite{santos00,yi01}. This is in contrast to the case of purely isotropic
interactions, for which the aspect ratio of the condensate matches that of
the trap.  By adopting an elongated aspect ratio along the direction of
polarization, the condensate achieves a lower energy by placing more dipoles
end-to-end in which configuration they are attractive, and reducing the
number of repulsive side-by-side interactions. When
$\varepsilon_{\mathrm{dd}} $ exceeds a certain value, which in general
depends on the aspect ratio the trap as well as the total number of atoms
$N$, mean-field theory predicts that the condensate becomes unstable to
collapse \cite{goral00,santos00,yi01,lushnikov,goral2002b}.  The partially
attractive nature of the dipole-dipole interaction is also responsible for
introducing a curious `roton' minimum \cite{santos03,odell03} into the
Bogoliubov excitation spectrum of a uniform dipolar BEC, reminiscent of that
found in the excitation spectrum of liquid helium II. It is possible that
this roton minimum is indicative of an instability towards the formation of
a density wave \cite{pitaevskii84,giovanazzi04}.

The fact that the interaction between dipoles aligned by an external field
along the axis of symmetry of a cylindrical trap are on average repulsive in
an oblate (pancake shaped) BEC, but, conversely, are on average attractive
in a prolate (cigar shaped) BEC, means that the sign of the dipolar
mean-field energy can be controlled via the aspect ratio of the trap.  In
this paper we are interested in calculating the effect that this change of
aspect ratio of the trap has upon the vortex state in a trapped dipolar BEC. The case of a vortex in
a trapped BEC with repulsive short-range interactions, as described by Eq.\
(\ref{eq:deltafninteraction}) with $a_{s}>0$, has been extensively studied
both theoretically, e.g.\ \cite{dalfovo96,lundh97,feder99}, and
experimentally, e.g.\ \cite{matthews99,madison00,raman01}.  Attractive
short-range interactions have, on the other hand, received comparatively
less attention, although a rotating BEC with attractive interactions is of
considerable fundamental interest \cite{wilkin98}.
In particular, when the
interactions are attractive it is intuitively plausible that vortex formation will be
suppressed because it is energetically expensive to form the vortex core
since this requires moving atoms from the centre of the BEC where they interact with many other atoms and placing them
on the edges where the interact with fewer \cite{JMFGunn}. Instead, it has been proposed that the angular
momentum might be preferentially absorbed into other types of excitation,
such as centre of mass motion or shape oscillations, which cause much less
disturbance to the condensate density profile and internal correlations
\cite{wilkin98,pethick+smith}. We also note that a recent theoretical study of a two-dimensional BEC with purely attractive short-range interactions found a new class of vortices in the form of bright ring solitons \cite{carr06}. When it comes to dipolar interactions, which are partly attractive, partly repulsive, one might therefore anticipate rotational properties similar to those of standard BECs with either
attractive or repulsive interactions, depending upon whether the dipolar BEC is prolate or oblate, respectively. 

The latest experiments on chromium \cite{stuhler05b} have detected the first
evidence of dipolar interactions upon the dynamics of a BEC by observing the
ballistic expansion of the BEC when the trap is switched off. The results
are in good agreement with theoretical predictions \cite{giovanazzi03}
giving us some confidence in the theory of dipolar BECs which has been built
up over the last six years. In the following we will discuss a vortex in a
rotating dipolar BEC in the Thomas-Fermi (hydrodynamic) regime, which is defined
as being when the quantum zero-point kinetic energy due to the confinement by the trap
is negligible in comparison to the trapping and
interaction energies. This is the relevant regime when there are a large number of
atoms. We shall make extensive use of the exact
results for dipolar BECs in the Thomas-Fermi limit reported in references \cite{odell04} and \cite{eberlein05}.  The
key insight of those papers was that the dipolar mean-field potential that
the atomic dipoles feel, and which is non-local, i.e. depends on the entire
atom distribution, can be expressed in terms of derivatives of a scalar
potential that is a solution of the Poisson equation with the dipolar
density as a source. In other words, this scalar potential is formally
equivalent to the electrostatic potential one would get if the density
distribution were not one of dipoles but of charges, and thus standard
mathematical techniques known from electrostatics can be applied. Three
recent theory papers \cite{cooper05,zhang05,yi06} have dealt with the structure of vortex lattices in
dipolar BECs, but our aim here is to investigate the differences
between dipolar and non-dipolar condensates in the case of a single
vortex. Our main concern will be the calculation of the critical rotation frequency $\Omega_{c}$ necessary to make a single vortex energetically favorable over the ground state. Note that reference \cite{yi06} also deals with the single vortex case, but their focus was somewhat different. One of their principle results was that in the case of dipolar interactions plus \emph{attractive} contact interactions (negative scattering length) the vortices develop a `craterlike' structure which is probably connected with the density wave instability mentioned above. We shall come back to this point in Section \ref{sec:energyfunctional}, however we shall not explicitly consider negative scattering lengths here because we wish to use some exact results that only hold in the Thomas-Fermi limit. A dipolar BEC with attractive short-range interactions is unstable to collapse in the Thomas-Fermi limit and so we limit ourselves to $a_{s}>0$.

\section{Energy functional for a dipolar BEC with a vortex} 
\label{sec:energyfunctional} 

Consider a condensate at $T=0$ with a wave function $\Psi(\br)$ normalized
to the total number of atoms $\int \vert \Psi(r) \vert^{2}
\romand^{3}r=N$. The general form of $\Psi(\br)$ when there is one singly
quantized vortex present can be expressed in cylindrical coordinates
$(\rho,\phi,z)$ as $\Psi(\br)=\vert \Psi(\rho,z) \vert \exp \mathrm{i} \phi$
\cite{pitaevskii+stringari}. This state carries an angular momentum $L_{z}=N
\hbar$. In a frame rotating at angular velocity $\Omega$ about the z-axis
the energy of the condensate becomes $E'=E-\Omega L_{z}$, where the angular
momentum $L_{z}$ and energy $E$ are those pertaining to the laboratory
frame. However, creating a vortex costs energy. Denoting the energy of the
BEC in its ground state without any vortex ($L_{z}=0$) by $E_{0}$ and the extra energy
needed to make a single vortex by $E_{v}$, we can write the energy of the vortex state in the rotating frame as 
\begin{equation}
E'=E_{0}+E_{v}-\Omega L_z\;.
\label{eq:E0+Ev}
\end{equation}
Thus we see that for a BEC at equilibrium in a trap rotating at angular
velocity $\Omega$ the vortex state becomes energetically favorable when
$\Omega$ exceeds a critical rotational velocity
\begin{equation}
\Omega_{c}=\frac{E_{v}}{N \hbar}.
\label{eq:omegacrit}
\end{equation} 
To obtain $\Omega_{c}$ we thus need to evaluate the extra energy associated
with the formation of the vortex.  The total energy functional for a trapped
dipolar BEC can be written as
\begin{equation}
E_{\mathrm{tot}}=E_{\mathrm{kinetic}}+E_{\mathrm{trap}}+E_{\mathrm{sw}}
+E_{\mathrm{dd}} \label{eq:Etot}
\end{equation}
where, in terms of the condensate wave function $\Psi(\br)$, 
\begin{equation}
E_{\mathrm{kinetic}}=-\frac{\hbar^{2}}{2m} \int \romand^{3}r  \  
\Psi^{*}(\br) \nabla^{2} \Psi(\br)
\label{eq:kedefn}
\end{equation}
is the kinetic energy, and
\begin{equation}
E_{\mathrm{trap}}=\frac{m}{2}  \omega_{x}^{2} \int \romand^{3}r \vert 
\Psi(\br) \vert^{2} \left[\rho^{2}+\gamma^{2} z^{2} \right]
\label{eq:trapdefn}
\end{equation}
is the energy due to the harmonic trap $V_{\mathrm{trap}}=(m/2) \omega_{x}^2
[ \rho^{2}+\gamma^{2}z^{2} ]$, where $\rho^{2}=x^{2}+y^{2}$ and 
\begin{equation}
\gamma
\equiv \omega_{z}/\omega_{x} 
\end{equation}
is the ratio of the trap frequencies. Note
that in this paper we shall assume that both the trap and the condensate are
cylindrically symmetric about the $\hat{z}$ direction, so in particular
$\omega_{x}=\omega_{y}$, and the external field responsible for aligning the
dipoles is  along the $\hat{z}$ axis.

The total mean-field interaction energy can to a good approximation
\cite{yi01} be written as the sum of two parts. The first is due to the
isotropic short-range interactions which give rise to pure s-wave scattering
\begin{equation}
E_{\mathrm{sw}}=\frac{g}{2} \int \romand^{3}r \vert \Psi(\br) \vert^{4} 
\label{eq:swdefn}
\end{equation}
and the second is due to dipole-dipole interactions
\begin{equation}
E_{\mathrm{dd}}=\frac{1}{2} \int \romand^{3}r  \  \romand^{3}r '  
\vert \Psi(\br) \vert^{2} 
U_{\mathrm{dd}}(\br-\br') \vert \Psi(\br') \vert^{2}.  \label{eq:Edd1}
\end{equation}
The long-range and anisotropic nature of the dipole-dipole interaction,
$U_{\mathrm{dd}}(\br)$, makes the calculation of $E_{\mathrm{dd}}$ a
non-trivial exercise.  However, in the Thomas-Fermi  regime some exact results (within mean-field theory) are available. In
particular, an exact solution of the vortex-free Thomas-Fermi problem for a
dipolar BEC in a harmonic trap yields a density profile which is an inverted
parabola \cite{odell04,eberlein05}. This solution is similar to the familiar
non-dipolar case \cite{pitaevskii+stringari}, except that the aspect ratio
of the cloud in the dipolar case is no longer identical to that of the trap
but is stretched along the direction of the polarizing field, as mentioned
above. By applying scaling transformations to this exact solution one can
also describe the lowest energy collective excitations of the condensate
\cite{odell04}, as well as the expansion dynamics when the trap is turned
off. In the Thomas-Fermi regime the critical ratio of the dipolar and
$s$-wave interactions above which instabilities occur is
$\varepsilon_{\mathrm{dd}}=1$ \cite{note1}. The occurrence of these
instabilities is indicated by the appearance of imaginary frequencies for
the collective excitations. Another particularly novel instability in a
trapped dipolar BEC, which originates from the long-range nature of the
interatomic interactions, is the appearance of a local minimum in the
mean-field potential \emph{outside} the condensate when
$\varepsilon_{\mathrm{dd}} > 1$. This stimulates the formation of a
`Saturn-ring' around the condensate as atoms tunnel into the minimum
\cite{eberlein05}.

As explained at length in references \cite{odell04} and \cite{eberlein05},
the dipole-dipole interaction can be re-expressed so as to take advantage of
the large technical machinery that exists to deal with electrostatic
interactions. In general one can write
\begin{equation}
E_{\mathrm{dd}}=\frac{1}{2} \int \romand^{3}r  \ n(\br) \Phi_{\mathrm{dd}}(\br)
\end{equation}
where $ \Phi_{\mathrm{dd}}(\br)$ is the non-local dipolar mean-field potential
\begin{equation}
\Phi_{\mathrm{dd}}(\br) \equiv \int \romand^{3}r'\
U_{\mathrm{dd}}(\br-\br') n(\br') \; . \label{eq:phidd}
\end{equation}
For dipoles aligned along $\hat{z}$ this dipolar mean-field potential can in
turn be expressed in terms of a fictious electrostatic potential $\phi(\br)$
\begin{equation}
\Phi_{\mathrm{dd}}(\br)  =  -C_{\mathrm{dd}}
\left(\frac{\partial^{2}}{\partial z^2} \phi(\br)  
+ \frac{n
(\br)}{3} \right)
\end{equation}
which is obtained from the ``charge'' distribution $n(\br)$ in the usual way
\begin{equation}
\phi(\br)  =  \frac{1}{4 \pi} \int \romand^{3} r' \frac{n(\br')}{\vert \br - 
\br' \vert}
\end{equation}
where, of course, $n(\br)$ and $\phi(\br)$ satisfy Poisson's equation
$\nabla^{2} \phi= - n (\br)$.  This formulation of the problem allows us to
immediately identify some generic features of dipolar gases. For example, if
$n(\br)$ and hence $\phi(\br)$ are uniform along the polarization direction
($\hat{z}$) then the non-local part of the dipolar interaction vanishes because of the $\partial^{2} \phi / \partial z^{2}$ operator. A
very prolate pencil-like condensate, i.e. one which has $R_{z} \gg R_{x}, R_{y}$ will
therefore be largely unaffected by the non-local part of the interaction
except perhaps at the very ends of the condensate if the density profile has a
large curvature there. Thus, in the very prolate (or entirely homogenous) case
the contribution of dipolar interactions to the mean-field potential reduces
to that of the local term $\Phi_{\mathrm{dd}}(\br) \rightarrow
-C_{\mathrm{dd}} n (\br)/3$ whose effect is then simply to modify the
magnitude of the local mean-field potential arising from the pure $s$-wave
interactions.
A related example to a very prolate BEC is a BEC with a generic smooth outer density profile but which is, however, penetrated by a pencil-like vortex or vortices: the dipolar potential would be expected  to peak around the ends of each vortex where the density profile is rapidly changing in the $z$-direction. One might conjecture that the curious `craterlike' density profile around the ends of vortices in dipolar BECs that has recently been seen in simulations \cite{yi06} is connected to this effect.

\begin{figure}[t]
\begin{center}
\centerline{\epsfig{figure=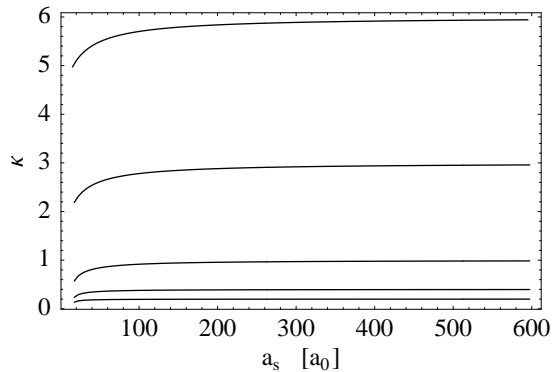,
  width= 8.0cm}}
\end{center}
\vspace*{-8mm} \caption{Aspect ratio $\kappa=R_{x}/R_{z}$ of a vortex-free dipolar BEC as a function of scattering length $a_{s}$ (measured in units of the Bohr radius $a_{0}$) in the Thomas-Fermi limit. In descending order, the curves are for traps with aspect ratios $\gamma= \omega_{z}/\omega_{x}=6,3,1,0.4,0.2$. When $a_{s} \rightarrow \infty$ the $s$-wave contact interactions dominate and the aspect ratio matches that of the trap, i.e.\ $\kappa=\gamma$, given by the right hand asymptote of each curve. When $a_{s} \rightarrow 0$ the dipolar interactions dominate and their magnetostrictive effect reduces $\kappa$ below the pure $s$-wave value. In the limit $a_{s} \rightarrow 0$ a dipolar BEC will collapse towards a chain of end-to-end dipoles and mean-field theory will breakdown. In this figure $C_{\mathrm{dd}}= \mu_{0} (6 \mu_{\mathrm{B}})^2$ and the minimum value of scattering length shown is $a_{s}=17.5 a_{0}$.}
\label{fig:kappaall}
\end{figure}

Let us consider the case of a parabolic density profile
of the form
\begin{equation}
 n_{\mathrm{bg}}(\br)=n_{0}\left(1-\frac{\rho^{2}}{R_{x}^{2}}
 -\frac{z^{2}}{R_{z}^{2}}\right)\;. \label{eq:nbg} 
\end{equation} 
As stated above, a parabolic density profile is an exact solution of the
hydrodynamic equations for a non-rotating BEC in the Thomas-Fermi limit even
in the presence of dipolar interactions. Here the subscript ``bg'' stands for ``background'', i.e. for a condensate
without a vortex but on top of which we will superimpose a vortex later on.
One finds that the dipolar mean-field potential inside a condensate with the parabolic density profile (\ref{eq:nbg}) is
given by \cite{odell04,eberlein05}
\begin{eqnarray}
\Phi_{\mathrm{dd}}^{\mathrm{bg}} (\rho,z)= \frac{ n_0
C_{\mathrm{dd}}}{3}\left[\frac{\rho^2}{ R_{x}^2}-\frac{2 z^2}{
R_{z}^2}-f \left( \kappa \right)\left(1-\frac{3}{2}\frac{\rho^2 -
2 z^2}{ R_{x}^2- R_{z}^2}\right) \right].
\label{phiddinside}
\end{eqnarray}
In this expression
\begin{equation}
f(\kappa) \equiv \frac{2+\kappa^{2}[4-3
\Xi(\kappa)]}{2(1-\kappa^{2})}
\label{eq:f}
\end{equation}
is a function of $\kappa$ which is the aspect ratio of the BEC,
\begin{equation}
\kappa \equiv R_{x}/R_{z}
\end{equation}
and monotonically
decreases from its value of $f=1$ at $\kappa=0$, passing through zero at
$\kappa=1$, tending towards $f= -2$ as $\kappa \rightarrow \infty$. The
function $\Xi(\kappa)$ upon which $f(\kappa)$ depends is itself dependent
upon whether the condensate density profile is prolate, in which case
\begin{equation}
\Xi \equiv  \frac{1}{\sqrt{1-\kappa^{2}}}\ln\frac{1
+\sqrt{1-\kappa^{2}}}{1-\sqrt{1-\kappa^{2}}} \quad \mathrm{for}
\quad \kappa<1 \quad \mathrm{(prolate)}
\end{equation}
or oblate
\begin{equation}
\Xi  \equiv  \frac{2}{\sqrt{\kappa^{2}-1}}\arctan
\sqrt{\kappa^{2}-1} \quad \mathrm{for} \quad \kappa>1 \quad
\mathrm{(oblate)}\;.
\end{equation}
The value of the aspect ratio $\kappa$
is given by the solution of a transcendental equation \cite{yi01,giovanazzi03,odell04,eberlein05}
\begin{equation}
3 \kappa^{2} \varepsilon_{\mathrm{dd}} \left[
\left(\frac{\gamma^{2}}{2}+1
\right)\frac{f(\kappa)}{1-\kappa^{2}}-1 \right] + \left(
\varepsilon_{\mathrm{dd}}-1 \right)
\left(\kappa^{2}-\gamma^{2}\right)  =0 .  \label{eq:transcendental}
\end{equation}
Note that according to Eq.\ (\ref{phiddinside}), the dipolar mean-field
potential $\Phi_{\mathrm{dd}}^{\mathrm{bg}} (\rho,z)$ is in general
saddle-shaped \cite{stuhler05b}, and thus inherits the anisotropic partially
attractive/partially repulsive character of the dipolar interactions. The saddle shaped mean-field potential causes an elongation of the BEC along the polarization direction which can be viewed as type of magnetostriction. This is illustrated for five different trap aspect ratios in Fig.\ \ref{fig:kappaall}. The
total energy ($E_{\mathrm{trap}}+E_{\mathrm{sw}}+E_{\mathrm{dd}}$)
associated with the vortex-free Thomas-Fermi solution (\ref{eq:nbg}) can be
calculated to be \cite{eberlein05}
\begin{equation}
E_{0}= \frac{N}{14} m \omega_{x}^{2} R_{x}^{2}
\left(2+\frac{\gamma^{2}}{\kappa^{2}} \right) + \frac{15}{28 \pi}
\frac{N^{2}g}{R_{x}^{2}R_{z}}\left[1-
\varepsilon_{\mathrm{dd}}f(\kappa) \right].
\label{eq:energyfunctional}
\end{equation}
The
quantity $E_{0}$ is the ground state energy referred in Eq.\
(\ref{eq:E0+Ev}).

Having now summarized what is already known concerning the mean-field
potential inside a non-rotating dipolar BEC, let us turn to the form of the
density profile of a rotating BEC with a vortex.  We shall adopt the following variational ansatz for the density profile of an $N$-atom condensate with a single vortex
\begin{eqnarray}
n(\br) & \equiv & \vert \Psi(\br)\vert^{2}  \nonumber \\
& = & n_{0}\left(1-\frac{\rho^{2}}{R_{x}^{2}}-\frac{z^{2}}{R_{z}^{2}}\right) 
\left(1-\frac{\beta^{2}}{\rho^{2}+\beta^{2}} \right) \label{eq:ansatz} \\
& = & n_{\mathrm{bg}}(\br)+ n_{\mathrm{v}}(\br)  \label{eq:ansatz2} 
\end{eqnarray}
where we have observed that the ansatz can be written as the sum of two
terms: the background Thomas-Fermi parabolic profile $n_{\mathrm{bg}}(\br)$
as already given in Equation (\ref{eq:nbg}) and the vortex profile
$n_{\mathrm{v}}(\br)$
\begin{equation}
n_{\mathrm{v}}(\br)  =  - n_{0}\frac{\beta^{2}}{\rho^{2}+\beta^{2}}
\left(1-\frac{\rho^{2}}{R_{x}^{2}} -\frac{z^{2}}{R_{z}^{2}}\right).
\end{equation}
Ansatz (\ref{eq:ansatz}) assumes that the vortex, whose core size is
parameterized by $\beta$, is superimposed on a background parabolic density
profile with radii $R_{x}=R_{y}$ and $R_{z}$.  Note that the variational ansatz
(\ref{eq:ansatz}) has the correct $\rho^{2\ell}$ dependence of the density
as $\rho \rightarrow 0$ (with $\ell=1$), which it must have in order to
satisfy the Gross-Pitaevskii equation for a vortex of $\ell$ circulation
quanta \cite{pethick+smith}. It has also got the correct asymptotic form (in
terms of even powers of $1/\rho$) that the solution must have for
$\rho\rightarrow\infty$. The central density $n_{0}$ is fixed by
normalization to be
\begin{eqnarray}
n_{0} & = & \frac{15 N}{8 \pi R_{x}^{2}R_{z} } \bigg/ \bigg( 1
+\frac{20}{3} \bar{\beta}^{2}+5 \bar{\beta}^{4}
\nonumber \\ && -5 \bar{\beta}^{2} \left(1+\bar{\beta}^{2}\right)^{3/2} 
\mathrm{arctanh} [1/\sqrt{1+\bar{\beta}^{2}}] \bigg) 
\label{eq:centraldensity}
\end{eqnarray}
where 
\begin{equation}
\bar{\beta} \equiv \beta/R_{x}
\end{equation} 
is the ratio of the vortex core size to the
transverse radius of the BEC.

We proceed by minimizing
$E_{\mathrm{tot}}$ as a function of the three variational parameters
$R_{x}$, $R_{z}$ and $\beta$.  The terms $E_{\mathrm{kinetic}}$,
$E_{\mathrm{trap}}$ and $E_{\mathrm{sw}}$ can all be calculated analytically
in a straightforward albeit laborious manner. The results are presented in
Appendix A.  To evaluate the dipole-dipole energy functional we begin from
Equation (\ref{eq:Edd1}) and substitute in the density ansatz
(\ref{eq:ansatz2})
\begin{eqnarray}
E_{\mathrm{dd}}  & = &  \frac{1}{2}  \int \romand^{3}r  \  \romand^{3}r '   
\; n_{\mathrm{bg}}(\br) U_{\mathrm{dd}}(\br-\br')  n_{\mathrm{bg}}(\br')
\nonumber\\
&& + \frac{1}{2}  \int \romand^{3}r  \  \romand^{3}r '\; n_{\mathrm{bg}}(\br) 
U_{\mathrm{dd}}(\br-\br')  n_{\mathrm{v}}(\br') \nonumber \\
&& + \frac{1}{2}  \int \romand^{3}r  \  \romand^{3}r '\; n_{\mathrm{v}}(\br)
U_{\mathrm{dd}}(\br-\br')  n_{\mathrm{bg}}(\br') \nonumber\\
&&+ \frac{1}{2}  \int 
\romand^{3}r  \  \romand^{3}r '\; n_{\mathrm{v}}(\br)
U_{\mathrm{dd}}(\br-\br')  
n_{\mathrm{v}}(\br') .\nonumber
\end{eqnarray}
Noting that the two cross terms between the vortex and background densities
are identical, i.e.\ the integral is invariant under exchange of the
coordinates $\br$ and $\br'$, the dipolar energy functional can be written
\begin{eqnarray}
E_{\mathrm{dd}} & = &   \frac{1}{2}  \int \romand^{3}r  \  \romand^{3}r '   
\; n_{\mathrm{bg}}(\br) U_{\mathrm{dd}}(\br-\br')  n_{\mathrm{bg}}(\br')
\nonumber\\ &&+    
\int \romand^{3}r  \  \romand^{3}r '\:   n_{\mathrm{v}}(\br)
U_{\mathrm{dd}}(\br-\br')  
n_{\mathrm{bg}}(\br') \nonumber \\  & & +   \frac{1}{2}  \int 
\romand^{3}r  \  \romand^{3}r ' \;  n_{\mathrm{v}}(\br)
U_{\mathrm{dd}}(\br-\br')  
n_{\mathrm{v}}(\br').\nonumber 
\end{eqnarray}
We shall now restrict ourselves to situations where the size of the vortex
core is much smaller than the radius of the condensate, i.e.\ $\bar{\beta}
\ll 1$. This is consistent with the spirit of the Thomas-Fermi approximation
for the background parabolic envelope, and should hold for condensates
containing a large number of atoms, providing the system does not become
very prolate. To this end we shall drop the vortex-vortex part of the
dipolar energy functional since this has an extra factor of $\bar{\beta}
^{2}$ in comparison to the cross term. The remaining two terms can be
expressed as
\begin{eqnarray}
E_{\mathrm{dd}}& \approx&  \frac{1}{2}  \int \romand^{3}r  \  
\romand^{3}r'\;   
n_{\mathrm{bg}}(\br) U_{\mathrm{dd}}(\br-\br')  n_{\mathrm{bg}}(\br') 
\nonumber\\&&+\int 
\romand^{3}r  \  \romand^{3}r ' \;  n_{\mathrm{v}}(\br)
U_{\mathrm{dd}}(\br-\br')  
n_{\mathrm{bg}}(\br') \nonumber \\  & = &  
\frac{1}{2} \int    \romand^{3}r  \;  n_{\mathrm{bg}}(\br) 
\Phi_{\mathrm{dd}}^{\mathrm{bg}}(\br) +   \int    \romand^{3}r  \;  
n_{\mathrm{v}}(\br) \Phi_{\mathrm{dd}}^{\mathrm{bg}}(\br) \ .
\end{eqnarray}
The first term is just the dipolar energy of the condensate without a
vortex, whilst the second gives the dipolar `interaction' between the vortex
and the background density profile. Both terms have been reduced to single
integrals of their respective density profiles over the known quadratic
function $\Phi_{\mathrm{dd}}^{\mathrm{bg}}(\rho, z)$, as given by Equation
(\ref{phiddinside}), which is the mean-field potential due to the background
inverted parabola density profile. These integrals can therefore be
evaluated relatively easily and the results are given in the Appendix.

\section{Results of minimization of the energy functional}
\label{sec:results}

\subsection{Oblate trap}

To illustrate the oblate case we shall consider 150,000 $^{52}$Cr atoms in a
trap having frequencies $\omega_{x}= \omega_{y}=2 \pi \times 200$ rad/s,
$\omega_{z}=2 \pi \times 1000$ rad/s, so that $\gamma=5$.  The harmonic
oscillator length of the trap along the x-direction is then
$a_{\mathrm{ho}}= \sqrt{\hbar/m \omega_{x}}=0.986 \mu$m. We numerically
minimize the total energy functional $E_{\mathrm{tot}}$ with respect to
\{$\beta$, $R_x$, $\kappa$\} using the \textit{Mathematica} routine
\textit{NMinimize}. In order to bring out the dependence of the various
quantities upon the relative strength of the two types of interactions
($s$-wave and dipolar) we plot the results as a function of the $s$-wave
scattering length $a_s$, since this quantity can be controlled in an
experiment \cite{werner05}. As $a_{s} \rightarrow 0$ the dipolar
interactions dominate, whereas when $a_{s} \rightarrow \infty$ the dipolar
interactions become insignificant. The magnitude of the magnetic dipole
interaction is here assumed to be fixed at the appropriate value for
$^{52}$Cr, namely $C_{\mathrm{dd}}=\mu_{0} (6\mu_{B})^2$. 
 Note that in the case of dipoles
induced by electric fields one has $ C_{\mathrm{dd}}=E^{2}
\alpha^{2}/\epsilon_{0}$ \cite{marinescu98}, where $\alpha$ is the 
polarizability and $E$ is the electric field strength, so the size of the
electric dipolar interaction can be directly controlled via the magnitude of
$E$. In fact the magnitude of even magnetic dipolar interactions can be
controlled independently of the $s$-wave interactions by rotating the
magnetic field \cite{giovanazzi2002b}, but we presume that the
experimentally easiest option is to use a Feshbach resonance to adjust $a_{s}$ since only the magnitude of the magnetic field needs to be tuned.

\begin{figure}[t]
\begin{center}
\centerline{\epsfig{figure=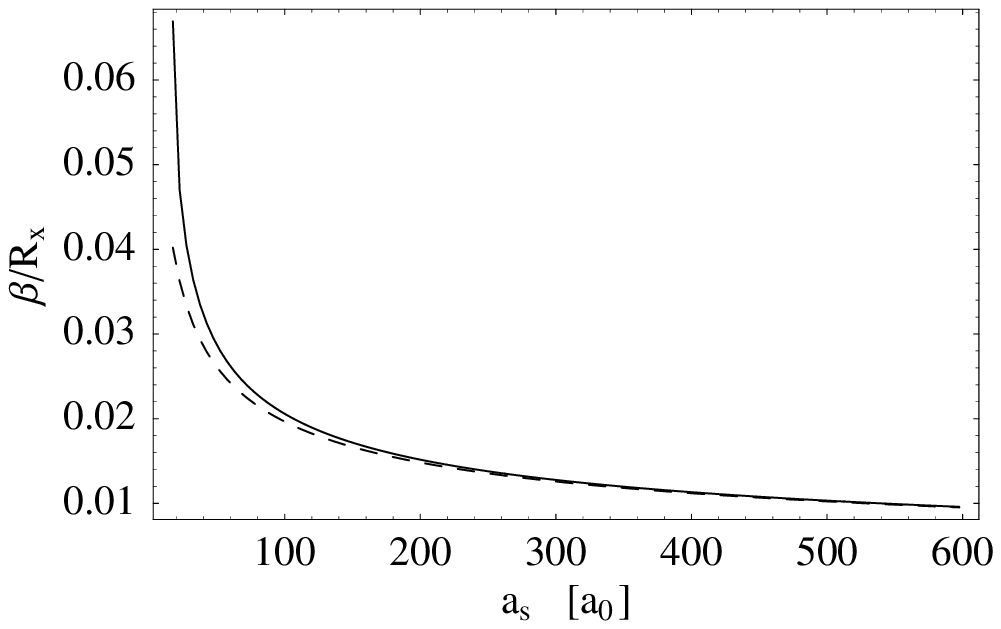,
  width= 8.0cm}}
\end{center}
\vspace*{-8mm} \caption{The ratio of the vortex core size $\beta$ to the 
radial size $R_x$ of a condensate in an oblate trap ($\gamma=5$).  Solid
curve: both $s$-wave and dipolar interactions. Dashed curve: $s$-wave
only. }
\label{fig:betapic1}
\end{figure}

In all the calculations depicted in the
following figures we limited the minimum value of the scattering length to be 
$17.5 a_0$, where  $a_0$ is the Bohr radius, corresponding to $\varepsilon_{\mathrm{dd}}=0.87$. This is because as $a_{s} \rightarrow 0$ the approximation we made in the calculation of the energy functional that $\bar{\beta}=\beta/R_{x}$ be small begins to breakdown (see Figures \ref{fig:betapic1} and \ref{fig:betapic1pro}).
Furthermore, in the Thomas-Fermi approximation used here collapse instabilities can occur when  $\varepsilon_{\mathrm{dd}}> 1$. Indeed, as mentioned previously, because of these collapse instabilities one can argue that the Thomas-Fermi regime does not exist when $\varepsilon_{\mathrm{dd}}> 1$, see \cite{eberlein05}.
 The maximum value of scattering length was set at $600 a_0$,  corresponding to $\varepsilon_{\mathrm{dd}}=0.03$.
 
\begin{figure}[t]
\begin{center}
\centerline{\epsfig{figure=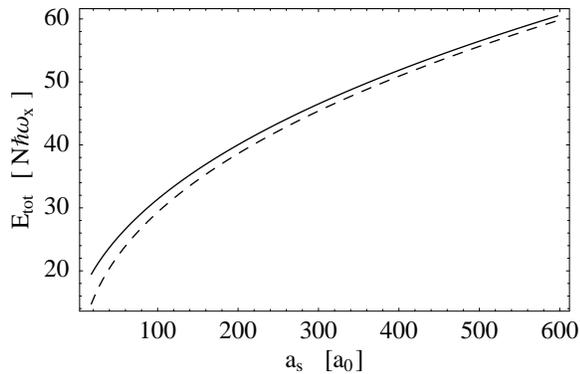,
  width= 8.0cm}}
\end{center}
\vspace*{-8mm} \caption{Total energy (in the Thomas-Fermi approximation) of 
a dipolar BEC with a vortex in an oblate trap ($\gamma=5$).  Solid curve:
both $s$-wave and dipolar interactions. Dashed curve: $s$-wave only. The
$s$-wave scattering length $a_{s}$ is measured in units of the Bohr radius
$a_{0}$. The energy $E_{\mathrm{tot}}$ is measured in units of the radial
harmonic trapping energy of N atoms. }
\label{fig:etot1}
\end{figure}

\begin{figure}[t]
\begin{center}
\centerline{\epsfig{figure=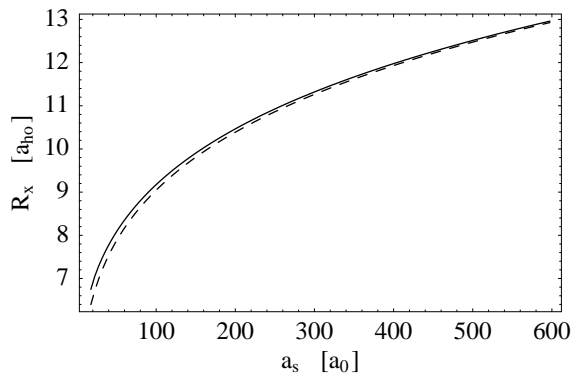,
  width= 8.0cm}}
\end{center}
\vspace*{-8mm} \caption{The radial size $R_{x}$ of a condensate with a 
vortex in an oblate trap ($\gamma=5$). Solid curve: both $s$-wave and
dipolar interactions. Dashed curve: $s$-wave only. $R_x$ is measured in
units of the radial harmonic oscillator length $a_{\mathrm{ho}}$ of the
trap. }
\label{fig:Rxpic1}
\end{figure}

\begin{figure}[t]
\begin{center}
\centerline{\epsfig{figure=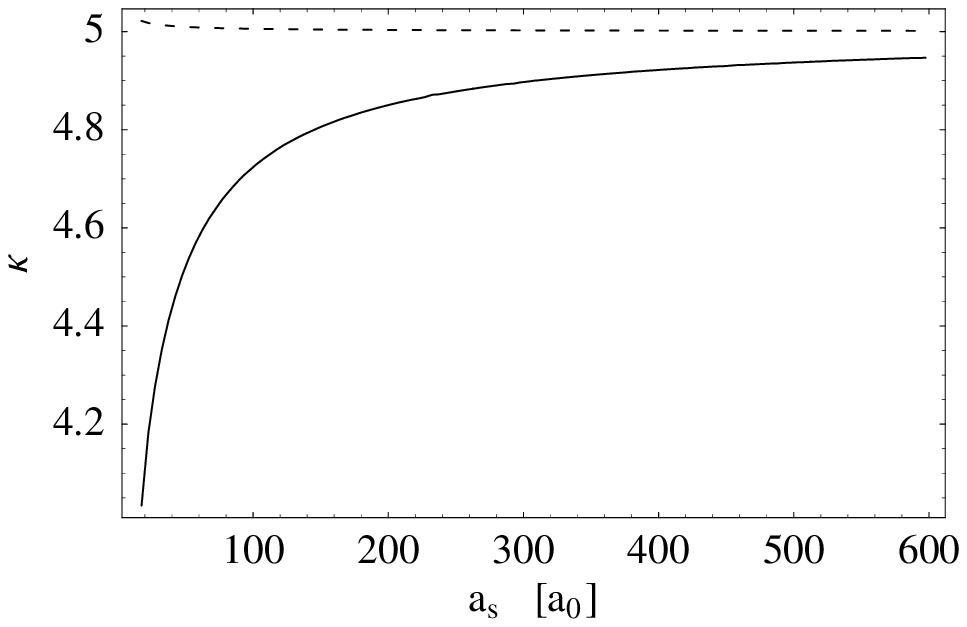,
  width= 8.0cm}}
\end{center}
\vspace*{-8mm} \caption{The aspect ratio $\kappa$ of a condensate with a 
vortex in an oblate trap ($\gamma=5$). Solid curve: both $s$-wave and
dipolar interactions. Dashed curve: $s$-wave only.  }
\label{fig:kappapic1}
\end{figure}
\begin{figure}[t]
\begin{center}
\centerline{\epsfig{figure=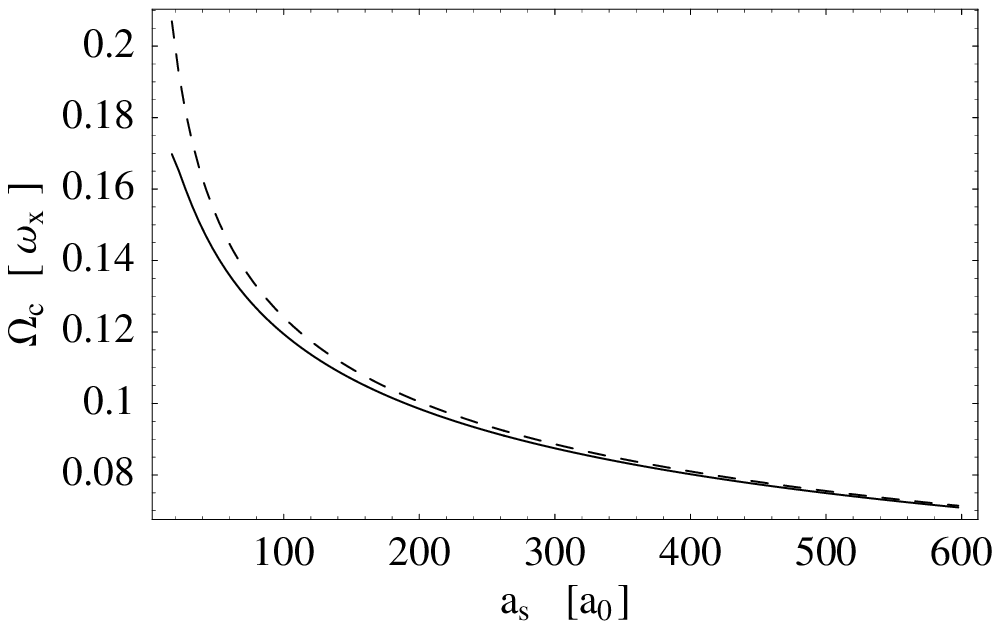,
  width= 8.0cm}}
\end{center}
\vspace*{-8mm} \caption{The critical angular velocity $\Omega_{c}$ above 
which a vortex state is energetically favorable in an oblate trap
($\gamma=5$) . Solid curve: both $s$-wave and dipolar interactions. Dashed
curve: $s$-wave only. These curves were calculated using Eq.\
(\ref{eq:omegacrit}). $\Omega_{c}$ is measured in units of the radial
harmonic oscillator angular frequency $\omega_{x}$ of the trap. }
\label{fig:omegacrit}
\end{figure}

\begin{itemize}
  \item Fig.\ \ref{fig:etot1} shows the total energy $E_{\mathrm{tot}}$, as defined
by Eq.\ (\ref{eq:Etot}), of a condensate with a vortex in an oblate trap. In an oblate BEC the dipolar interactions are predominantly repulsive, which raises the energy
relative to the $s$-wave only case.
  \item Fig.\ \ref{fig:Rxpic1} plots the radial size $R_{x}$ of a condensate with a
vortex. In a strongly oblate BEC such as this one the majority of atoms interact via the radially repulsive part of the dipolar interaction which consequently  increases $R_{x}$ slightly above the pure $s$-wave value.
  \item Fig.\ \ref{fig:betapic1} shows $\bar{\beta}=\beta/R_{x}$, the ratio of the
vortex core size to the radial size of the condensate. Note that this ratio remains small over the chosen range of parameters, which, as mentioned above, is necessary for the self-consistency of the calculation.  In an oblate trap we see that the effect
of the dipolar interactions is to increase $\bar{\beta}$ beyond that found in the pure $s$-wave case.
 \item Fig.\ \ref{fig:kappapic1} depicts the aspect ratio $\kappa=R_{x}/R_{z}$ of a condensate with a vortex. The magnetostriction that reduces $\kappa$
by elongating the condensate is clearly visible.
\item Fig.\ \ref{fig:omegacrit} gives the critical angular velocity $\Omega_{c}$ of the condensate above which it is energetically favorable to form a vortex.  We see that the effect of dipolar interactions in an oblate trap is to \emph{decrease} $\Omega_{c}$. 
\end{itemize}

\subsection{Prolate trap}

We now turn to the case of a prolate trap for which the dipolar interactions
are predominantly attractive.  We choose a trap which has the inverse aspect
ratio to the previous oblate case, namely with frequencies $\omega_{x}=
\omega_{y}=2 \pi \times 200$ rad/s as before, so that the harmonic
oscillator length of the trap along the x-direction remains the same, but
with $\omega_{z}=2 \pi \times 40$ rad/s, so that $\gamma=0.2$.

\begin{figure}[t]
\begin{center}
\centerline{\epsfig{figure=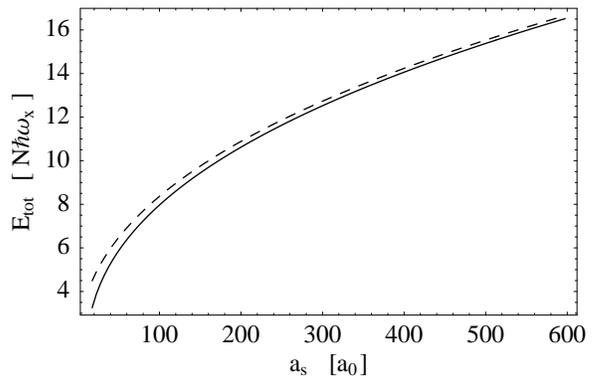,
  width= 8.0cm}}
\end{center}
\vspace*{-8mm} \caption{Total energy (in the Thomas-Fermi approximation) 
of a dipolar BEC with a vortex  in a prolate trap ($\gamma=0.2$).    
Solid curve: both $s$-wave and dipolar interactions. Dashed curve:  
$s$-wave only. }
\label{fig:etot1pro}
\end{figure}

\begin{figure}[t]
\begin{center}
\centerline{\epsfig{figure=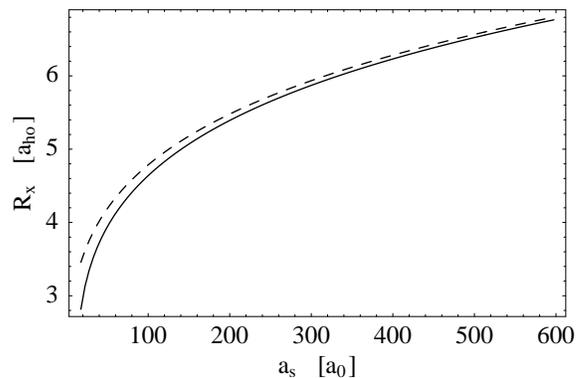,
  width= 8.0cm}}
\end{center}
\vspace*{-8mm} \caption{The radial size $R_{x}$ of a condensate with a 
vortex in a prolate trap ($\gamma=0.2$). Solid curve: both $s$-wave and
dipolar interactions. Dashed curve: $s$-wave only.}
\label{fig:Rxpic1pro}
\end{figure}

\begin{figure}[t]
\begin{center}
\centerline{\epsfig{figure=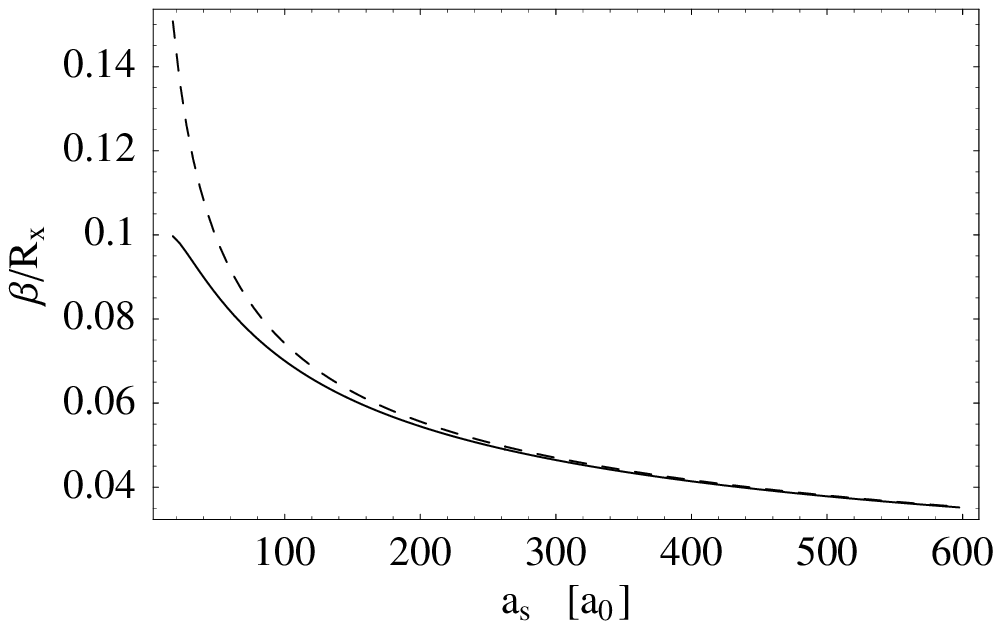,
  width= 8.0cm}}
\end{center}
\vspace*{-8mm} \caption{The ratio of the vortex core size $\beta$ to the 
radial size $R_x$ of a condensate in an prolate trap ($\gamma=0.2$).  Solid
curve: both $s$-wave and dipolar interactions. Dashed curve: $s$-wave
only. }
\label{fig:betapic1pro}
\end{figure}

\begin{figure}[t]
\begin{center}
\centerline{\epsfig{figure=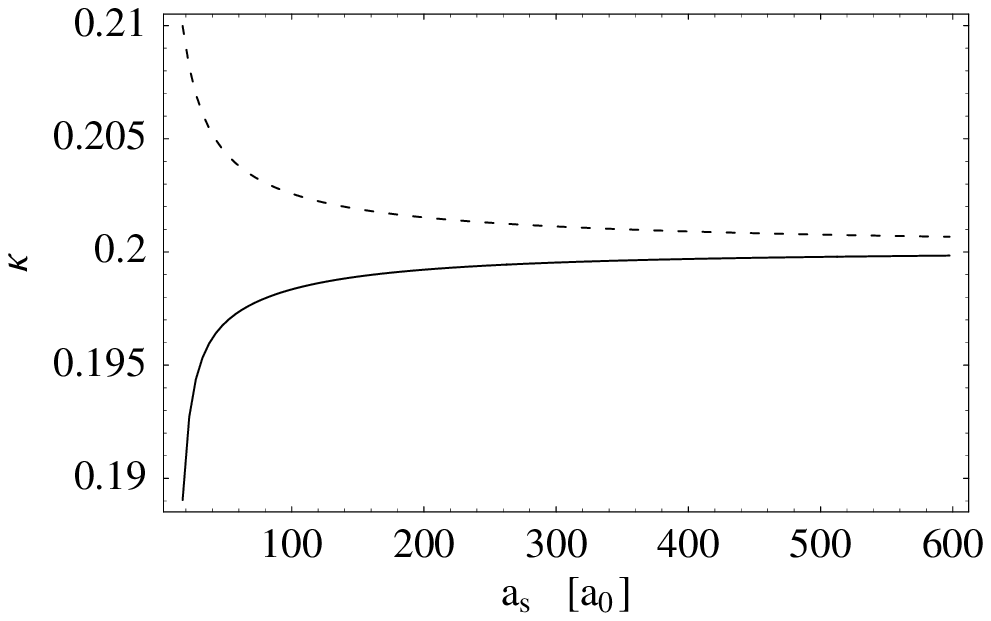,
  width= 8.0cm}}
\end{center}
\vspace*{-8mm} \caption{The aspect ratio $\kappa$ of a condensate with a 
vortex in a prolate trap ($\gamma=0.2$). Solid curve: both $s$-wave and
dipolar interactions. Dashed curve: $s$-wave only.  }
\label{fig:kappapic1pro}
\end{figure}

\begin{figure}[t]
\begin{center}
\centerline{\epsfig{figure=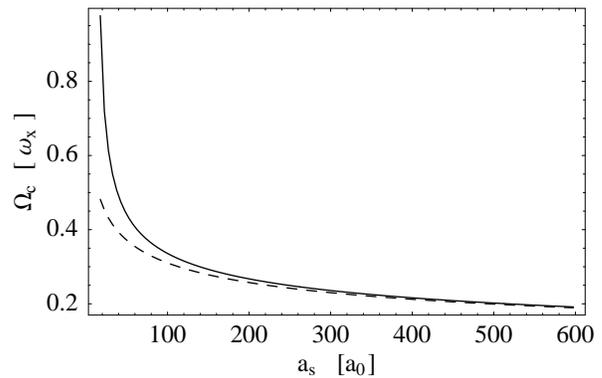,
  width= 8.0cm}}
\end{center}
\vspace*{-8mm} \caption{The critical angular velocity $\Omega_{c}$ above 
which a vortex state is energetically favorable in a prolate trap
($\gamma=0.2$) . Solid curve: both $s$-wave and dipolar interactions. Dashed
curve: $s$-wave only. }
\label{fig:omegacritpro}
\end{figure}

\begin{itemize}
\item Fig.\ \ref{fig:etot1pro} depicts the total energy $E_{\mathrm{tot}}$ of the 
vortex state in an prolate trap. The effect of the mainly attractive dipolar
interactions is to shift the energy downwards, oppositely to the case of an 
oblate trap.
\item Fig.\ \ref{fig:Rxpic1pro} plots the radial size, $R_{x}$, of a dipolar condensate with a
vortex in a prolate trap. $R_{x}$ is smaller than in the pure $s$-wave case because of the re-distribution of atoms by the anisotropic interactions.
\item Fig.\ \ref{fig:betapic1pro} shows that the ratio of the vortex core size
to the radial size of the condensate, $\bar{\beta}=\beta/R_{x}$. Note that this ratio remains small over the chosen range of parameters, which is necessary for the self-consistency of the calculation, although not as small as in the oblate case.
Notice also that in comparison the pure $s$-wave case  $\bar{\beta}$
 is smaller in the presence of dipolar interactions in a prolate trap, and this trend is
opposite to that found in an oblate trap. 
\item Fig.\ \ref{fig:kappapic1pro} gives the aspect ratio $\kappa=R_{x}/R_{z}$ of a condensate with a vortex in a
prolate trap. Due to the magnetostriction $\kappa$ is smaller than in the pure $s$-wave case. This is the same behavior as was found in an oblate trap.
\item Fig.\ \ref{fig:omegacritpro} shows that, in contrast to the
oblate case, the dipolar interactions in a prolate trap \emph{increase} the value of the
critical angular velocity $\Omega_{c}$ of the condensate above which it is energetically
favorable to form a vortex.  
\end{itemize}

\section{An explicit expression for $\Omega_{c}$}
\label{sec:analyticsol}

The effect that $\Omega_{c}$ is decreased in an oblate trap and increased in a prolate trap is the main result of this paper. Let us see if we can interpret this physically.
The value of $\Omega_{c}$ shown in Figures \ref{fig:omegacrit} and  \ref{fig:omegacritpro} for the purely $s$-wave
interactions (dashed curve) agrees very well with the analytic formula valid
in the Thomas-Fermi limit \cite{lundh97}
\begin{equation}
\Omega_{c}=\frac{5}{2} \frac{\hbar}{mR_{x}^{2}} \ln \frac{0.67 R_{x}}{\xi_{s}}.
\label{eq:lundhSol}
\end{equation}
This is derived by integrating the kinetic energy density  $n(\rho,z) v_{s}^{2}(\rho)/2$,  arising from the superfluid flow $v_{s} = (\hbar/m)1/\rho$ around the vortex, over the profile of the condensate. The lower limit of the integral is set by the vortex core size which is given by the  healing length $\xi_{s}$ 
\begin{equation}
\xi_{s}=\frac{1}{\sqrt{8 \pi n_{0}a_{s}}}.
\label{eq:healinglength}
\end{equation}
The relevant density appearing in this expression is taken to be the density at the centre of the trap $n_{0}$ in the absence of a vortex, which in the Thomas-Fermi limit  is given by 
\begin{equation}
n_{0}=\frac{15 N}{8 \pi} \frac{1}{R_{z}R_{x}^{2}}.
\end{equation}
In the Thomas-Fermi regime the radii of a condensate with $s$-wave interactions obey
\cite{pitaevskii+stringari}
\begin{equation}
R_{i}=\overline{a}_{\mathrm{ho}} \left(\frac{15Na_{s}}{\overline{a}_{\mathrm{ho}}} \right)^{1/5} 
\frac{\overline{\omega}_{\mathrm{ho}}}{\omega_{i}}
\label{eq:SWradius}
\end{equation} 
where $\overline{\omega}_{\mathrm{ho}}=(\omega_{x}\omega_{y}\omega_{z})^{1/3}$ and
$\overline{a}_{\mathrm{ho}}=\sqrt{\hbar/(m\overline{\omega}_{\mathrm{ho}})}$. 
Combining all these quantities together one finds that as $a_{s} \rightarrow 0$ the
critical rotation frequency diverges as $\Omega_{c} \propto a_{s}^{-2/5} \log a_{s}$,  although, of course, eventually the neglected quantum pressure term will prevent this divergence.

Now consider the dipolar case. A formula for $\Omega_{c}$ closely resembling (\ref{eq:lundhSol}) should still apply. Indeed, the inverted-parabola Thomas-Fermi density profile is common to both the $s$-wave and dipolar cases and it is this that gives rise to the specific 5/2 and 0.67 numerical factors appearing in (\ref{eq:lundhSol}). The one difference we might expect concerns the lower limit of the integration which is set by the vortex core size.
In the presence of dipolar interactions we can no longer assert that the core size is solely determined by $\xi_{s}$ since this makes no reference to dipolar interactions. Indeed, we might expect that as the scattering length vanishes it is replaced by an equivalent length scale set by the dipolar interactions of the form $a_{d} \equiv (C_{\mathrm{dd}}/3) m/(4 \pi \hbar^{2})$. However, expression (\ref{eq:lundhSol}) only has logarithmic accuracy \cite{StatPhysII} and is relatively insensitive to the lower cutoff $\xi_{s}$ of the kinetic energy integral appearing inside the logarithm. The dominant change in  (\ref{eq:lundhSol}) due to the presence of dipolar interactions is therefore likely to be in the radial size $R_{x}$. This makes one wonder whether the same expression (\ref{eq:lundhSol}) for $\Omega_{c}$ approximately holds for dipolar BECs if the radius $R_{x}$ is modified to include the effect of the dipole-dipole interactions but the changes in the healing length are ignored? 
The answer to this question is yes \cite{santosprivate} as we shall now show. 

\begin{figure}[t]
\begin{center}
\centerline{\epsfig{figure=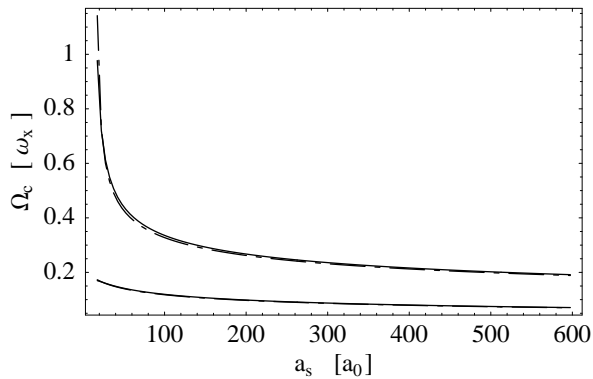,
  width= 8.0cm}}
\end{center}
\vspace*{-8mm} \caption{Comparison of the critical angular velocity $\Omega_{c}$ calculated by the methods of Section \ref{sec:results} (solid curves) with the prediction of the analytic formula Eq.\ (\ref{eq:lundhSol}) where the condensate radius $R_{x}$ is calculated for a dipolar BEC in the Thomas-Fermi limit with no vortex (dash-dot curves), see text for details. The upper two curves are for the prolate case, $\gamma=0.2$, and the lower two curves are for the oblate case, $\gamma=5$, as before. Note that the match is so good that it is hard to discern the difference between the solid and the dash-dot curves, except for very small $a_{s}$ in the prolate case. 
  }
\label{fig:santosapprox}
\end{figure}

The Thomas-Fermi radii of a vortex-free BEC cylindrically symmetric in the $x-y$ plane (dipoles aligned along $z$) with both contact and dipolar interactions are \cite{odell04,eberlein05}
\begin{eqnarray}
R_{x}=R_{y}=\left[\frac{15 g N \kappa}{4 \pi m \omega_{x}^{2}}
\left\{1+  \varepsilon_{\mathrm{dd}} \left( \frac{3}{2}
\frac{\kappa^{2} f(\kappa)}{1-\kappa^{2}}-1 \right)
 \right\} \right]^{1/5} \label{eq:Rxsol}
\end{eqnarray}
and $R_{z}=R_{x}/\kappa$. The function $f(\kappa)$ appearing in this expression is given by Eq.\ (\ref{eq:f}) and the value of $\kappa$ is determined by the transcendental Eq.\ (\ref{eq:transcendental}).
In Fig.\ (\ref{fig:santosapprox}) we compare the value of $\Omega_{c}$ calculated by the energy minimization method presented in Section \ref{sec:results} with the value of $\Omega_{c}$ obtained from the explicit expression (\ref{eq:lundhSol}) where $R_{x}$
is given by  (\ref{eq:Rxsol}).
The agreement is strikingly good, with the exception of very small $a_{s}$ in the prolate case. In fact, in view of the closeness of the match one is tempted to conclude that whilst the long-range dipolar interactions strongly influence the boundary of the condensate and hence the large scales represented by $R_{x}$, the shorter range van der Waals interactions which set the scale for $a_{s}$ and hence $\xi_{s}$ continue to dominate the shorter range physics setting the size of the vortex core, except when $a_{s}$ becomes very small indeed.

\section{Comparison of vortices with other types of angular motion}

We have seen in previous sections that for a dipolar BEC in a rotating
prolate trap, the formation of a vortex is increasingly energetically
suppressed as the dipolar interactions grow in strength. Indeed, once
$\Omega_{c} \ge \omega_{x}$ the vortex would be impossible to realize, in a
harmonic trap at least, because the trap can no longer counteract the
centrifugal force and the BEC would fly apart (neglecting, of course, the
intriguing possibility that the attractive nature of the mean-field
interactions might be capable of holding the condensate together for
rotational frequencies $\Omega$ greater than $\omega_{x}$). The question
then arises, how does a prolate dipolar BEC rotate? One possibility is for
the condensate to generate other types of excitation that carry angular
momentum, such as quadrupole shape oscillations which have an azimuthal angular momentum projection $m=2$. In the frame of reference
rotating with the trap a quadrupole oscillation appears as a stationary
distortion of the density profile in the $x-y$ plane
\cite{recati01,sinha01} and it is generated through a dynamical instability which has a threshold frequency $\Omega_{\mathrm{dyn}}$ that in the pure $s$-wave case is typically considerably higher than $\Omega_{c}$ (in the case of an axisymmetric trap $\Omega_{\mathrm{dyn}}=\omega_{x}/\sqrt{2}$). Dynamical instabilities corresponding to shape oscillations play an important role in the \emph{dynamics} of how a vortex actually enters the condensate, a subject we have not touched upon since we have only considered the energetics rather than the dynamics of vortex formation. In the case of pure $s$-wave condensates it is an experimental fact \cite{madison00} that in order to generate vortices the trap must be rotated at a frequency considerably exceeding the condition of energetic stability given by Eq.\ (\ref{eq:lundhSol}) and instead corresponding  closely to $\Omega_{\mathrm{dyn}}$. 

The dynamical instability of a rotating dipolar BEC to shape oscillations has recently been investigated \cite{vanBijnen06}, and it was found that
dipolar interactions always lower the threshold rotational frequency $\Omega_{\mathrm{dyn}}$ at which the dynamical instability occurs, in both the prolate \emph{and} oblate cases.  It is interesting to note that in the pure $s$-wave case the threshold frequency for the instability is \emph{independent} of interaction strength (even though the instability relies on the existence of two-body interactions \cite{pitaevskii+stringari}) but, conversely, in the dipolar case is strongly affected by the magnitude of the interactions (but perhaps, based on the discussion of Section \ref{sec:analyticsol}, the effect of dipolar interactions is simply through the degree to which these change the shape and aspect ratio of the condensate). One might tentatively conjecture, therefore, that as the critical rotation frequency for energetic favorability of the vortex state goes up, and the critical rotation frequency for the dynamical instability towards shape oscillations goes down, shape oscillations will be formed that will not go on to form vortices.

Let us now also briefly
consider another possibility, namely oscillations of the centre of mass
(CM). In a harmonic trap CM oscillations decouple from internal excitations
provided the interactions are pairwise, and therefore take place at the trap
frequency.  Like low-lying shape oscillations, CM oscillations also preserve
the parabolic density profile and so might be energetically preferable to a
vortex in the presence of attractive interactions.  Consider a BEC executing
a CM oscillation of the form
\begin{equation}
\br_{cm}(t)=r_{0} \cos (\omega_{x} t) \hat{\mathbf{x}} + r_{0} \sin 
(\omega_{x} t) \hat{\mathbf{y}}
\label{eq:com}
\end{equation}
where $\br_{cm}(t)$ is the position of centre of mass of the BEC, which
behaves like a classical particle of mass $Nm$ with potential energy $NM
\omega_{x}^{2} r_{cm}^{2}/2$.  This circular motion has an angular momentum
$L_{cm}=Nm \omega_{x} r_{0}^{2}$, with the actual value of $L_{cm}$ being
determined by the radius $r_{0}$.  Note that (\ref{eq:com}) describes
dynamics obeying the superfluid irrotationality condition ($\nabla \times
v_{s}=0$) as may be verified by observing that the wave function which is an
exact solution of the time-dependent Gross-Pitaevskii equation describing
this type of motion is of the form \cite{pitaevskii+stringari}
\begin{eqnarray}
\Psi(\br,t) & = &\exp [-\mathrm{i} \mu t/\hbar] 
\exp \left\{\mathrm{i} \alpha(t)[x-A(t)/2]/\hbar\right\} 
\nonumber\\ && \times\exp \left\{\mathrm{i} \beta(t) 
[y-B(t)/2]/\hbar\right\}\nonumber \\
&& \times \Psi_{0}(x-A(t),y-B(t),z)
\end{eqnarray}
where $\Psi_{0}(x,y,z)$ is a stationary state satisfying the time-independent
Gross-Pitaevskii equation, and the parameters $\{A(t),\alpha(t); B(t),
\beta(t)\}$ obey $\alpha=m \dot{A}$, $Am \omega_{x}^{2}=-\dot{\alpha}$ and
$\beta=m \dot{B}$, $Bm \omega_{x}^{2}=-\dot{\beta}$.

The energy associated with the type of CM oscillations given by
(\ref{eq:com}) is $E_{cm}=Nm \omega_{x}^{2} r_{0}^{2}$. In order to make a
comparison between the energy of the vortex state and the CM motion we set
$L_{cm}=N \hbar$, so that the two have the same angular momentum. Then
$r_{0}^{2}=\hbar/(m \omega_{x})$, i.e. the radius of motion is equal to the
trap oscillator length, and so $E_{cm}^{L=N \hbar}=N\hbar \omega_{x}$.
Noting that the ground state energy, $E_{0}$, is the same for both the
vortex and CM motion, we can therefore compare $E_{cm}$ with $E_{v}=N \hbar
\Omega_{c}$.  In order to be competitive with a vortex, CM motion with
$L_{cm}=N \hbar$ would therefore require $\Omega_{c} \ge \omega_{c}$, but
this is exactly the rotational speeds for which the harmonic trap no longer
confines the atoms and so we surmise that CM motion is not energetically
favorable over a vortex state in a harmonic trap. This does not preclude, however, that centre
of mass motion takes place in preference to other types motion for angular
momenta $L < N \hbar$. Also, perhaps centre of mass motion can be favored in non-harmonic traps.

\section{Conclusions}
The question of how quantum fluids rotate is a
fascinating one. The versatility of atomic BECs means that 
this problem can now be studied in systems with repulsive or attractive interactions, as well as the case of dipole-dipole interactions which are partially repulsive partially attractive and long range. In this paper we have used an approach based on an analogy to electrostatics that allows the explicit calculation of the anisotropic mean-field potential inside a dipolar BEC.  
This approach gives quite general insight and in particular shows that the long-range part of the interaction is especially important in places where the density profile has a large curvature in the direction of polarization, such as the ends of vortices.
Numerical minimization of the total energy functional calculated by this method in the Thomas-Fermi regime indicates that, in comparison to the case of pure $s$-wave contact interactions, dipolar interactions lower the critical rotation frequency $\Omega_{c}$ of a BEC necessary to make a vortex energetically favorable in an oblate trap and raise the critical rotation frequency in a prolate trap. The same results can also be accurately reproduced using the analytic formula Eq.\ (\ref{eq:lundhSol}) well known in the usual case of $s$-wave contact interactions providing the modification of the radius $R_{x}$ of the BEC due to dipolar interactions is accounted for using Eq.\ (\ref{eq:Rxsol}). The analytic formula allows one to attribute the principal change in $\Omega_{c}$ caused by the dipolar interactions to changes in $R_{x}$, i.e. large scale changes in the overall shape of the condensate, rather than changes on the much smaller scale of the healing length which determines the size of the vortex core.   Finally, we have also discussed angular momentum carrying shape oscillations as well as centre of mass oscillations as competitors to vortices in rotating prolate dipolar BECs, and tentatively conclude that under certain circumstances shape oscillations such as the quadrupole oscillation may be preferred.

\begin{acknowledgments}
We would like to thank S.\ Giovanazzi and A.M.\ Martin for many discussions. We are indebted to J.M.F.\ Gunn for first suggesting to us the problem discussed in this paper and to
L. Santos for proposing we try the explicit solution described in Sec \ref{sec:analyticsol}. We gratefully acknowledge financial
support from the Engineering and Physical Sciences Research Council (EPSRC)
of the UK in the form of a postdoctoral fellowship in theoretical physics as well as QIP-IRC postdoctoral funding (D.O'D.) and from the Royal Society (C.E.).
\end{acknowledgments}

\appendix
\section{The energy functionals}

In this appendix we write down the explicit expressions for the functionals
giving the kinetic, trapping, $s$-wave scattering, and dipolar interaction
energy of the condensate, as defined by Equations (\ref{eq:kedefn}),
(\ref{eq:trapdefn}), (\ref{eq:swdefn}), and (\ref{eq:Edd1}),
respectively. Beginning with the kinetic energy, we first note that in
cylindrical coordinates the gradient is given by
\begin{equation}
\mathbf{\nabla}= \hat{\mathbf{e}}_{\rho} \frac{\partial }{\partial \rho}+ 
\hat{\mathbf{e}}_{\phi}\frac{1}{\rho}\frac{\partial }{\partial \phi} +  
\hat{\mathbf{e}}_{z} \frac{\partial }{\partial z}
\end{equation}
which acts upon the wave function given by the square root of the density 
ansatz (\ref{eq:ansatz})
\begin{equation}
 \Psi (\rho,\phi, z)=\sqrt{n_{0}} \sqrt{1-\frac{\rho^{2}}{R_{x}^{2}} 
-\frac{z^{2}}{R_{z}^{2}}} \sqrt{ 1-\frac{\beta^{2}}{\rho^{2}+\beta^{2}} }
\   \mathrm{e}^{\mathrm{i} \phi}.
\end{equation}
The phase $\phi$ accounts for the superflow around the vortex in the usual
way, viz.\ $\mathbf{v}_{s}=(\hbar/m) \pmb{\nabla} \phi$
\cite{pitaevskii+stringari}.  However, the Thomas-Fermi approximation
corresponds to neglecting the kinetic energy arising from the curvature of
the slowly varying condensate background envelope
$\sqrt{1-\rho^{2}/R_{x}^{2} -z^{2}/R_{z}^{2}}$. We shall consequently drop
all terms which originate from the gradient of this term. On the other hand,
it is essential to retain the terms arising from the gradient of the vortex
part of the wave function $\sqrt{ 1-\beta^{2}/(\rho^{2}+\beta^{2}) }$ which
varies rapidly in the $ \hat{\mathbf{e}}_{\rho} $ direction, as well as the
superflow term. Under these approximations we find the kinetic energy to be
\begin{eqnarray}
E_{\mathrm{kinetic}}&=& \frac{\hbar^{2}}{2m} \frac{n_{0} \pi R_{z}}{3}
\bigg\{-11\bar{\beta}^2 - \frac{26}{3}  \\&&+ \frac{11 
\bar{\beta}^4 + 16 
\bar{\beta}^2+ 8}{\sqrt{1 + \bar{\beta}^2}} \mathrm{arctanh} \left[1/\sqrt{1
+ \bar{\beta}^2} \right] \bigg\}\nonumber
\end{eqnarray}
where $n_{0}$ is the number density at the centre of the trap, as given by 
Equation (\ref{eq:centraldensity}).  

Using the density ansatz (\ref{eq:ansatz}) the trapping and $s$-wave
scattering energies are straightforwardly evaluated to be
\begin{eqnarray}
E_{\mathrm{trap}}&=& \frac{2\pi}{3}n_{0}m \omega_{x}^{2} 
\frac{R_{x}^{5}}{\kappa} \Bigg\{ \frac{4}{35} - \frac{2}{5}\bar{\beta}^2
-\frac{8}{3}\bar{\beta}^4 -2\bar{\beta}^6 \nonumber\\&&
+\frac{2\gamma^2}{5\kappa^2}\left(\frac{1}{7}+ \frac{23}{15}\bar{\beta}^2
+ \frac{7}{3}\bar{\beta}^4 + \bar{\beta}^6\right)\nonumber\\
&&+2\bar{\beta}^2\left[\bar{\beta}^2 -\frac{\gamma^2}{5\kappa^2}
\left(1+\bar{\beta}^2\right) \right]\left(1+\bar{\beta}^2\right)^{3/2}
\nonumber\\&&\hspace*{18mm}\times
\mathrm{arctanh}\left(1/\sqrt{1+ \bar{\beta}^2}\right)\Bigg\}
\end{eqnarray}
and
\begin{eqnarray}
E_{\mathrm{sw}}&=&\frac{8\pi}{15}g n_{0}^{2} \frac{R_{x}^{3}}{\kappa}
\left[ \frac{2}{7} + \frac{107}{15}\bar{\beta}^2 +16\bar{\beta}^4
  +9\bar{\beta}^6\right.\nonumber\\&&
-\bar{\beta}^2\left(4+13\bar{\beta}^2+9\bar{\beta}^4\right)
\sqrt{1+\bar{\beta}^2}\nonumber\\&&\left.\hspace*{16mm}\times
\mathrm{arctanh}\left(1/\sqrt{1+ \bar{\beta}^2}\right)\right].
\end{eqnarray}

The dipolar energy functional is more difficult to calculate since the 
dipolar mean-field potential $ \Phi_{\mathrm{dd}}(\br)$ is non-local. 
In the text we argued that 
\begin{equation} E_{\mathrm{dd}} \approx \frac{1}{2} \int    \romand^{3}r  
\  n_{\mathrm{bg}}(\br) \Phi_{\mathrm{dd}}^{\mathrm{bg}}(\br) +   \int    
\romand^{3}r  \  n_{\mathrm{v}}(\br) \Phi_{\mathrm{dd}}^{\mathrm{bg}}(\br)
\end{equation}
where $ \Phi_{\mathrm{dd}}^{\mathrm{bg}}(\br)$ is given by the quadratic 
function Equation (\ref{phiddinside}). These two integrals can be evaluated 
explicitly and the results are
\begin{equation}
\frac{1}{2} \int    \romand^{3}r  \  n_{\mathrm{bg}}(\br) 
\Phi_{\mathrm{dd}}^{\mathrm{bg}}(\br)  = -\frac{16 \pi g n_{0}^{2}}{105\kappa} 
\varepsilon_{\mathrm{dd}} R_{x}^{3} f(\kappa) 
\end{equation}
\begin{widetext}
and 
\begin{eqnarray}
\int    \romand^{3}r  \  n_{\mathrm{v}}(\br) 
\Phi_{\mathrm{dd}}^{\mathrm{bg}}(\br) & = &
-\frac{4\pi}{15}n_{0}^{2}g\varepsilon_{\mathrm{dd}}
\frac{R_{x}^{3}}{\kappa}\bar{\beta}^2 \bigg\{
\frac{122}{15}+\frac{68}{3} \bar{\beta}^2 +14\bar{\beta}^4\nonumber\\&&
+\frac{f(\kappa)}{15(\kappa^2-1)}\left[ -62 +245\kappa^2
+30\bar{\beta}^2\left(2+15\kappa^2\right) 
+45\bar{\beta}^4\left(2+5\kappa^2\right) \right] \\
&&-\left(1+\bar{\beta}^2\right)^{3/2}
\mathrm{arctanh}\left(1/\sqrt{1+ \bar{\beta}^2}\right)
\left[4+14\bar{\beta}^2 +\frac{f(\kappa)}{\kappa^2-1} \left(
-4+10\kappa^2+6\bar{\beta}^2+15\kappa^2\bar{\beta}^2\right)
\right]\bigg\}\;.\nonumber
\end{eqnarray}
\end{widetext}


\begin{thebibliography}{99}
\bibitem{griesmaier05}
{A. Griesmaier, J. Werner, S. Hensler, J. Stuhler and T. Pfau,
Phys. Rev. Lett. \textbf{94}, 160401 (2005).}
\bibitem{pitaevskii+stringari}
{L. Pitaevskii and S. Stringari, \textit{Bose-Einstein Condensation} 
(Oxford, 2003).}
\bibitem{werner05}
{J. Werner, A. Griesmaier, S. Hensler,  J. Stuhler, T. Pfau, A. Simoni 
and E. Tiesinga, Phys. Rev. Lett. \textbf{94}, 183201 (2005).}
\bibitem{santos00}
{L. Santos, G.V. Shlyapnikov, P. Zoller, and M. Lewenstein, Phys.
Rev. Lett. \textbf{85}, 1791 (2000).}
\bibitem{yi01}
{S. Yi and L. You, Phys. Rev. A \textbf{63}, 053607 (2001).}
\bibitem{goral00}
{K. G{\'{o}}ral, K. Rz{\c{a}}{\.{z}}ewski, and T. Pfau, Phys. Rev.
A {\bf 61}, 051601(R) (2000); J.-P. Martikainen, M. Mackie, and K.-A.
Suominen, Phys. Rev. A \textbf{64}, 037601 (2001).}
\bibitem{goral2002b}
{K. G{\'{o}}ral and L. Santos, Phys. Rev. A \textbf{66}, 023613
(2002).}
\bibitem{lushnikov}
{P.M. Lushnikov, Phys. Rev. A \textbf{66}, 051601(R) (2002).}
\bibitem{santos03}
{L. Santos, G.V. Shlyapnikov, and M. Lewenstein, Phys. Rev. Lett.
\textbf{90}, 250403 (2003).}
\bibitem{odell03}
{D.H.J. O'Dell, S. Giovanazzi and G. Kurizki, Phys. Rev. Lett.
\textbf{90}, 110402 (2003).}
\bibitem{pitaevskii84}
{L. P. Pitaevskii, Zh. Eksp. Teor. Fiz. 39, 423 (1984)
[ , JETP Lett. 39, 511 (1984)].}
\bibitem{giovanazzi04}
{S. Giovanazzi and D. O'Dell, Eur.\ Phys.\ J.\ D.  \textbf{31}, 439 (2004).}
\bibitem{dalfovo96}
{F. Dalfovo and S. Stringari, Phys. Rev. A \textbf{53}, 2477 (1996).}
\bibitem{lundh97}
{E. Lundh, C.J. Pethick and H. Smith, Phys. Rev. A \textbf{55}, 2126 (1997).}
\bibitem{feder99}
{D.L. Feder, C.W. Clark, and B.I. Schneider, Phys. Rev. Lett. \textbf{82}, 
4956 (1999).}
\bibitem{matthews99}
{M.R. Matthews, B.P. Anderson, P.C. Haljan, D.S. Hall, C.E. Wieman, 
and E.A. Cornell, Phys. Rev. Lett. \textbf{83}, 2498 (1999).}
\bibitem{madison00}
{K.W. Madison, F. Chevy, W. Wohlleben and J. Dalibard, Phys. Rev. Lett. 
\textbf{84}, 806 (2000); K.W. Madison, F. Chevy, V. Bretin, and J. Dalibard, Phys. Rev. Lett. 
\textbf{86}, 4443 (2001).}
\bibitem{raman01}
{C. Raman, J.R. Abo-Shaeer, J.M. Vogels, K. Xu, and W. Ketterle, 
Phys. Rev. Lett. \textbf{87}, 210402 (2001).}
\bibitem{wilkin98}
{N.K. Wilkin, J.M.F. Gunn, and R.A. Smith, Phys. Rev. Lett. \textbf{80}, 
2265 (1998)}.
\bibitem{JMFGunn}
{J.M.F. Gunn, private communication.}
\bibitem{pethick+smith}
{C.J. Pethick and H. Smith, \emph{Bose-Einstein Condensation in Dilute Gases},
(Cambridge, 2002). See, in particular, Fig.\ 9.3.}
\bibitem{carr06}
{L.D. Carr and C.W. Clark, Phys. Rev. Lett. \textbf{97}, 010403 (2006).}
\bibitem{stuhler05b}
{J. Stuhler, A. Griesmaier, T. Koch, M. Fattori, T. Pfau, 
S. Giovanazzi, P. Pedri and L. Santos,
Phys. Rev. Lett. \textbf{95}, 150406 (2005).}
\bibitem{giovanazzi03}
{S. Giovanazzi, A. G\"{o}rlitz and T. Pfau, J. Opt. B \textbf{5} S208 (2003).}
\bibitem{odell04}
{D.H.J. O'Dell, S. Giovanazzi and C. Eberlein, Phys. Rev. Lett.
\textbf{92}, 250401 (2004).}
\bibitem{eberlein05}
{C. Eberlein, S. Giovanazzi and D.H.J. O'Dell, Phys.\ Rev.\ A \textbf{71}, 
033618 (2005).}
\bibitem{cooper05}
{N.R. Cooper, E.H. Rezayi, and S.H. Simon, Phys. Rev. Lett. \textbf{95}, 
200402 (2005).}
\bibitem{zhang05}
{J. Zhang and H. Zhai, Phys. Rev. Lett. \textbf{95}, 200403 (2005).}
\bibitem{yi06}
{S. Yi and H. Pu, Phys. Rev. A \textbf{73}, 061602(R) (2006).}
\bibitem{note1}
{ This is most easily seen by comparing the Fourier transform of the dipolar 
interaction
(\ref{eq:staticdipdip}), i.e.\ $
\tilde{U}_{\mathrm{dd}}(\bk)= \frac{C_{\mathrm{dd}}}{3}\,
\hat{{\rm e}}_{i} \hat{{\rm e}}_{j} (3 \hat{k}_{i} \hat{k}_{j}
-\delta_{ij})$, with that of the contact interaction, i.e.\
$\tilde{U}(\bk)=g$. One immediately sees that $\varepsilon_{\mathrm{dd}}=1$
corresponds to the case when the isotropic term in the dipole-dipole
interaction is exactly equal and opposite to the contact interaction.}
\bibitem{marinescu98}
{M. Marinescu and L. You, Phys. Rev. Lett. \textbf{81}, 4596
(1998).}
\bibitem{giovanazzi2002b}
{S. Giovanazzi, A. G\"{o}rlitz and T. Pfau, Phys. Rev. Lett.
\textbf{89}, 130401 (2002).}
\bibitem{StatPhysII}
{E.M. Lifshitz and L.P. Pitaevskii, \textit{Statisitcal Physics Part 2} 
(Butterworth-Heinemann, Oxford, 1998).}
\bibitem{santosprivate}
{L. Santos, private communication.}
\bibitem{recati01}
{A. Recati, F. Zambelli and S. Stringari, Phys. Rev. Lett. \textbf{86}, 377 
(2001).}
\bibitem{sinha01}
{S. Sinha and Y. Castin, Phys. Rev. Lett. \textbf{87}, 190402 (2001).}
\bibitem{vanBijnen06}
{R.M.W. van Bijnen, D.H.J. O'Dell, N.G. Parker and A.M. Martin, 
cond-mat/0602572.}



\end{thebibliography}
\end{document}